\documentclass[prl,aps,twocolumn,superscriptaddress]{revtex4}
\usepackage{comment}
\usepackage{amssymb}
\usepackage{amsmath}
\usepackage[dvipsnames,usenames]{color}
\usepackage{graphicx,graphics}
\usepackage{amssymb}

\newcommand{\dis}{\displaystyle}

\begin{document}
\title{Enhanced squeezing with parity kicks}

\author{Xiao-Tong Ni}
\affiliation{The Key Laboratory of Atomic and Nanosciences, Ministry
of Education, Tsinghua University, Beijing 100084, China }

\author{Yu-xi Liu}
\affiliation{Institute of Microelectronics, Tsinghua
University,Beijing 100084, China} \affiliation{Tsinghua National
Laboratory for Information Science and Technology, Tsinghua
University, Beijing 100084, China}
\author{L C Kwek}
\affiliation{Center for Quantum Technologies, National University of
Singapore,2 Science Drive 3, Singapore 117542 and National Institute
of Education and Institute of Advanced Studies, Nanyang
Technological University, 1 Nanyang Walk, Singapore 637616}

\author{Xiang-Bin Wang}
  \email{xbwang@mail.tsinghua.edu.cn}
  \affiliation{The Key Laboratory of Atomic and Nanosciences, Ministry
of Education, Tsinghua University, Beijing 100084, China }
\affiliation{Tsinghua National Laboratory for Information Science
and Technology, Tsinghua University, Beijing 100084, China}
  \affiliation{Department of Physics, Tsinghua University, Beijing 100084, China}

\begin{abstract}
Using exponential quadratic operators, we present a general
framework for studying the exact dynamics of system-bath interaction
in which the Hamiltonian is described by the quadratic form of
bosonic operators. To demonstrate the versatility of the approach,
we study how the environment affects the squeezing of quadrature
components of the system. We further propose that the squeezing can
be enhanced when parity kicks are applied to the system.
\end{abstract}

\maketitle

{\em Introduction} -- Coupling between system and environment is
ubiquitous in all quantum processes (e.g.,  in quantum information
processing). Such coupling usually results in: (i) the energy decay
of the quantum system; (ii) the destruction of the relative phases
of several superposed quantum states, and thus the linear
superposition of several quantum states turn into a classical
mixture. However, the environment can also help us, for example,
entanglement between two systems can be generated via a common
environment~\cite{Paz2008}.

Although it seems impossible to model the environment exactly in
many cases, and thus difficult to obtain the exact dynamics of a
system-environment interaction,  quantitative analysis based on
approximate description of the environment is needed in many cases,
for instance,  the analysis of decoherence suppressing
methods~\cite{Misra1977,Erez2008,Vitali1999,Viola2005,Kaveh2007,Ponte2004,Kofman2004,
Ulrike2000,Wu2002}; the discussion of the entanglement of two
systems coupled to the environment; and the study of the quantum
dissipation of systems~\cite{Leggett1987,Weiss}. An extensively
adopted approach to model the environment, which is also called a
reservoir or bath, is to introduce a set of harmonic oscillators
with different frequencies. In this case, the interaction between
the system and the environment is modeled by coupling the system to
these harmonic oscillators through an appropriate interaction
Hamiltonian. Several methods have been proposed to study the
coupling between the system and a set of harmonic oscillators. In
quantum optics (e.g.,
Refs.~\cite{Scully1997,Louisell,Gardiner1991}), a quite often used
method to analyze Markovian process is either a master equation or a
Langevin equation. Another method is the path integral approach
~\cite{Feynman1963} which was extensively developed in
Refs.~\cite{Leggett1983,Leggett1987}, but, this method is very
complicated.  Furthermore, different approximations are used in all
of these methods to make the problem tractable for either analytical
or numerical calculations.

In this paper, we introduce a new method to calculate the evolution
of the bosonic system, coupled to the environment. The total
Hamiltonian is described by a quadratic form of the bosonic
operators. Our method is based on some properties of exponential
quadratic operators. As shown in below, this new method provides a
feasible way to calculate the effect of the environment on system.
As an example, we apply our method to study the environment effect
on the generation of squeezed states. Moreover, we also use our
method to study the system-environment interaction when the parity
kicks are applied to the system. We find that the parity kicks can
help us to obtain a better squeezing.

 {\em Exponential quadratic operators.---}For a set of annihilation
operator $a_{i} (1\leq i \leq n)$, exponential quadratic
operators(EQO) (see \cite{eqo1,eqo2,eqo3}) are expressions of the
form
\begin{equation}
Q=e^{\sum_{i,j}(c_{ij}a_{i}a_{j}+d_{ij}a_{i}a_{j}^{\dag}+e_{ij}a_{i}^{\dag}a_{j}^{\dag})}.
\end{equation}
Here $[a_{i},a_{i}^{\dag}]=1$.  The above equation can also be
written in the following way $Q=e^{\frac{1}{2}\Lambda^{T}R\Lambda}$
in which
$\Lambda^{T}=(a_{1}^{\dag},a_{2}^{\dag},\cdots,a_{n}^{\dag},a_{1},a_{2},\cdots,a_{n})$
and $R$ is a symmetric matrix. If we define $\displaystyle S=
\begin{pmatrix}
  0 & I \\
  -I & 0 \\
\end{pmatrix}$,
then we have
\begin{equation}\label{thm}
    Q\Lambda^{T}Q^{-1}=\Lambda^{T}e^{-RS},
\end{equation}
where the multiplication in $e^{Q}\Lambda^{T}e^{-Q}$ is understood
to act on each term of $\Lambda$.

{\em Coupling between oscillator and reservoir.---} Consider a
system comprising of a harmonic oscillator with annihilation
operator $a$, and a reservoir consisting of a set of oscillators
with annihilation operator $b_k$ for each mode. The Hamiltonian of
system-reservoir is described by
\begin{equation}\label{h1}
    H=\hbar \omega
a^{\dag}a+\sum_{k}\hbar\omega_{k}b_k^{\dag}b_k+\hbar\sum_{k}\gamma_{k}(ab_k^{\dag}+b_{k}a^{\dag}),
\end{equation}
where the first, second, and third terms are  the system, reservoir,
and system-reservoir interaction Hamiltonians, respectively. Here,
$\gamma_{k}$ are the coefficients representing the coupling strength
between the system and the mode $k$ of reservoir - these coupling
constants are typically much smaller than the other frequencies in
the Hamiltonian. For simplicity,  but without loss of generality, we
regard these couplings as reals.

We calculate the evolution of $a$($a^{\dag}$) in the Heisenberg
picture by using equation (\ref{thm}). For $U=e^{-iHt/\hbar}$, putting
$\Lambda^{T}=(a^{\dag},b_{1}^{\dag},b_{2}^{\dag},\cdots,b_{n}^{\dag},a,b_{1},b_{2},\cdots,b_{n})$
and
$\displaystyle
R=\begin{pmatrix}
     & P \\
    P &  \\
  \end{pmatrix}
$
where
\begin{equation}
P=
    \begin{pmatrix}
      i\omega t & i\gamma_{1}t & i\gamma_{2}t & \cdots & i\gamma_{n}t \\
      i\gamma_{1}t & i\omega_{1}t &  &  &  \\
      i\gamma_{2}t &  & i\omega_{2}t &  &  \\
      \vdots &  &  & \ddots &  \\
      i\gamma_{n}t &  &  &  & i\omega_{n}t \\
    \end{pmatrix}
\end{equation}
Thus, to calculate the evolution of $a$ ($a^{\dag}$), we need only
to calculate the matrix $e^{-RS}$.

{\em Coupling between system and reservoir during a squeezing
process.---} To see the power of the technique, let us consider a
Hamiltonian for degenerate parametric amplification with a classical
pump under the influence of a reservoir in a squeezing process. The
Hamiltonian can be expressed as
\begin{equation}\label{totalsqueezehamilton}
\begin{split}
    H&=\hbar\omega a^{\dag}a+\frac{1}{2}i\hbar\epsilon[e^{2i\omega t}a^2-e^{-2i\omega t}(a^{\dag})^2]\\
    &+\sum_{j=1}^{n}\hbar
    \omega_{j}b_{j}^{\dag}b_{j}+\hbar(a^{\dag}\sum_{j=1}^{n}\gamma_{j}b_{j}+h.c.).
\end{split}
\end{equation}
In order to remove the time dependence in the Hamiltonian, we
transfer the Hamiltonian into a rotating reference frame with
$U=\exp(iH_{0}t/\hbar)$ with $\dis
H_{0}=\hbar\omega(a^{\dag}a+\sum_{j=1}^{n}b_{j}^{\dag}b_{j})$. Thus
in the rotating reference frame, the Hamiltonian in
Eq.~(\ref{totalsqueezehamilton}) becomes
\begin{equation}\label{hi}
\begin{split}
    H_{I}=&-\frac{1}{2}i\hbar\epsilon[(a^{\dag})^2-a^2] +
    \sum_{j=1}^{n}\hbar(\omega_{i}-\omega)b_{j}^{\dag}b_{j}\\
    &+\hbar(a^{\dag}\sum_{j=1}^{n}\gamma_{j}b_{j}+h.c.).
\end{split}
\end{equation}
Note that the first term is just the usual squeezing operator. We
can easily find the matrix R corresponding to $-iH_{I}t/\hbar$. By
analyzing $e^{-RS}$, numerically if necessarily, we obtain the
evolution of $a^{\dag}(t)$ and $a(t)$, and thus the solution of all
quantities associated with a squeezing process. The most important
one among them is
$\langle(\Delta(a(t)+a(t)^{\dag}))^2\rangle=\langle(\Delta
X)^2\rangle$.

{\em Parity kicks in the squeezing process.---} Using appropriate
time varying control fields, it is well known that one could
alleviate decoherence effects through a sequence of frequent parity
kicks. As in Ref.~\cite{Vitali1999}, we introduce an extra
Hamiltonian (in the rotating reference frame).
$$H_I^{\prime}=H_I+H_{kick}(t),$$
where $H_{kick}(t)=H_{kick}$ for $t_i\leq t \leq t_i+\tau$ and
$H_{kick}=0$ otherwise. We require $t_{i+1}-t_i=\tau_0$ for all $i$.
Moreover, we assume $\tau\ll\tau_0$ and $H_{kick}(t)$ is strong
enough during the kick periods that we can neglect the effect of
$H_I$, which is $\displaystyle e^{-iH_I \tau/\hbar}\approx
e^{-iH_{kick} \tau/\hbar}$ Under these conditions, we will model
parity kicks as unitary operators $P=e^{-iH_{kick} \tau/\hbar}$
acting on system at a set of time $t_i$. Since we want to eliminate
the influence of coupling between system and reservoir, we require
$P$ to have following properties $PH_{system}P=H_{system},$
$PH_{bath}P=H_{bath},$ and $PH_{int}P=-H_{int}.$ The three
Hamiltonians are defined in Eq.~\eqref{hi}. It is easy to verify
that $P=e^{-i\pi a^{\dag}a}$ satisfies above equations. Thus the
unitary operator corresponding to two such periods would be $
\displaystyle Pe^{-iH_I\tau_0/\hbar}Pe^{-iH_I\tau_0/\hbar}
=e^{-(i\tau_0/\hbar)(H_{system}+H_{bath}-H_{int})}e^{-(i\tau_0/\hbar)(H_{system}+H_{bath}+H_{int})}
\doteq Y. $ Intuitively, it shows the interaction between system and
reservoir of different periods cancel each other out. In fact, it
has been proved that when $\tau_0\rightarrow 0$ the system and the
reservoir are totally decoupled.

We use numerical computation to verify this effect in the squeezing
process. To this end we calculate the evolution of $a^{\dag}(t)$ and
$a(t)$: We have
$$a^{\dag}(2n\tau_0)=Y^{\dag n}a^{\dag}Y^n.$$
To use the EQO method shown in Eq.~\eqref{thm} to solve the above
expression, we note that if
$$e^{Y_1}\Lambda^{T}e^{-Y_1}=\Lambda^{T}P_1,$$
$$e^{Y_2}\Lambda^{T}e^{-Y_2}=\Lambda^{T}P_2,$$
then
$$e^{Y_2}e^{Y_1}\Lambda^{T}e^{-Y_1}e^{-Y_2}=\Lambda^{T}P_2P_1.$$
Thus, we know that we need only to calculate the $e^{-RS}$ in
$$Y^{\dag}\Lambda^{T}Y=\Lambda^{T}e^{-RS},$$
and $e^{-nRS}$ would be the desired transforming matrix. Again we
use the above property and see that we only need to calculate
\begin{equation}\label{banghint}
e^{(i\tau_0/\hbar)(H_{system}+H_{bath} \pm
H_{int})}\Lambda^{T}e^{-(i\tau_0/\hbar)(H_{system}+H_{bath} \pm
H_{int})}.
\end{equation}

\begin{figure}
  \includegraphics[width=8.8cm]{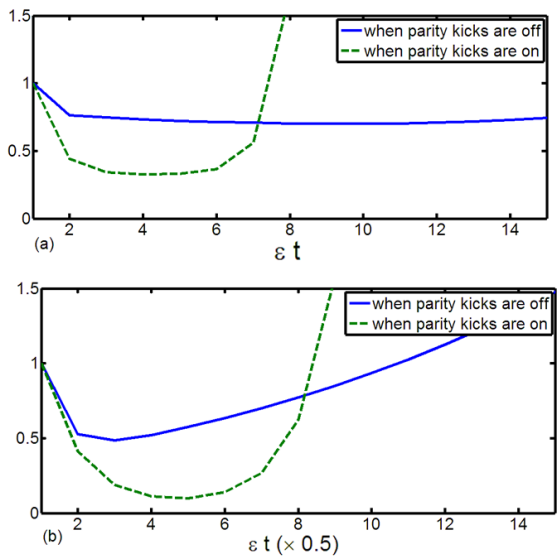}
  \caption{The variance $\langle(\Delta X)^2\rangle$ versus rescaled time $\
\epsilon t$ when parity kicks are on and off
  respectively. Figure (a) refers to a Lorentzian spectrum and figure (b) refers
  to an Ohmic spectrum.
  We find that a better squeezing  can be obtained when parity kicks are applied to the system.
  }\label{fig:paritykicks}
\end{figure}

For simplicity we consider the ground state situation. The procedure
is entirely general and applies for the case of $T>0$K. We compute
the variance $\langle(\Delta X)^2\rangle$ with two types of
coupling: namely, the Lorentzian spectrum and the ohmic spectrum.
For the Lorentzian spectrum
$\gamma_j=g(\omega_j)=\eta\Gamma/\sqrt{(\omega_{j}-\omega)^2+\Gamma^{2}},$
as an example of numerical calculations, we assume
$\Gamma=2\times10^9$Hz, $\eta=5\times 10^7$Hz, the squeezing
parameter $\epsilon=10^{8}$Hz and the kick period $\tau_0=1.67\times
10^{-9}$s. For the ohmic spectrum
$\gamma_j=g(\omega_j)=\sqrt{\xi\omega_j}e^{-\omega_j/\omega_c}$, we
assume $\xi=10^6$Hz, $\omega_c=10^9$Hz, the squeezing parameter
$\epsilon=7\times 10^{8}$Hz and the kick period $\tau_0=2.5\times
10^{-9}$s. For both spectrum, we assume the frequencies associated
with the system and reservoir to be $\omega=10^{9}$Hz and
$\omega_j=j\times10^{7}$Hz $(j=1,2,\ldots 200)$ respectively. With
these parameters, the variance $\langle(\Delta X)^2\rangle$ versus
rescaled time $\ \epsilon t$  is plotted in
Fig.~\ref{fig:paritykicks}, which shows that a better squeezing can
be obtained if parity kicks are applied to the system.

\begin{figure}
  \includegraphics[width=8.8cm]{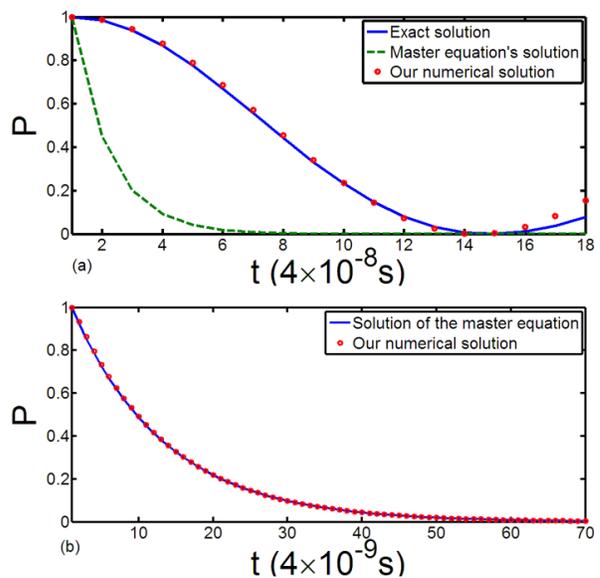}
  \caption{(a) The probability of the system being in state $|1\rangle$ with a Lorentzian spectrum. We can see that our numerical solution is close to the exact solution.
  On the other hand, we can see master equation is not valid for this situation.
  (We set the reservoir to have 200 equally distributed oscillators while getting the numerical solution of our method.)
  (b) The probability of the system being in state $|1\rangle$ when the coupling strengths $\gamma_k$ are constant.
  We can see that the lines of our numerical solution and master equation's solution coincide with each other.}\label{fig:compare}
\end{figure}

{\it Discussions.---} It is interesting to do a comparison between
several methods, including the widely used Markovian master equation
\cite{Louisell}. Under the Hamiltonian \eqref{h1} with a Lorentzian
spectrum as shown in Ref.~\cite{Liu2001}, we can obtain an exact
solution
\begin{equation}\label{lorentzexact}
\begin{split}
    a(t)=&[u(t)e^{-\Gamma t/2}a+\sum u_j(t)b_j]e^{-i\omega t}\\
    =&\{[\cos(\Theta t/2)+\frac{\Gamma}{\Theta}\sin(\Theta t/2)]+\sum u_j(t)b_j\}e^{-i\omega
    t}.
    \end{split}
\end{equation}
The constant $\Theta$ is given by $\Theta=\sqrt{4\pi \eta^2
D-\Gamma^2}$, where D is the density of reservoir modes and $u_j(t)$
are some complicated functions~\cite{Liu2001}. However, the master
equation, in the Markovian approximation, can be written as
\begin{equation}\label{markovianmaster}
\begin{split}
    \frac{\partial \rho}{\partial
    t}=&-i\omega[a^{\dag}a,\rho]+\frac{\lambda}{2}
    [2a\rho a^{\dag}-a^{\dag}a\rho-\rho a^{\dag}a]+\\
    &+\lambda
    \overline{n}[a^{\dag}\rho a+a\rho a^{\dag}-a^{\dag}a\rho-\rho
    aa^{\dag}],
\end{split}
\end{equation}
where $\lambda=2\pi D g(\omega)^2$ is a constant, which represents
the decay rate of the harmonic oscillator. For simplicity, we assume
the temperature of reservoir to be zero, and the initial state of
system to be $|1\rangle$. We then calculate the the probability
$P(t)=\langle 1 |tr_{R}(\rho(t))|1\rangle$, which can be used to
observe the decay of system. We can also compute $P(t)$ when the
coupling strengths $\gamma_k$ are constants. For example, we assume
that parameters of the Lorentzian spectrum in
Fig~\ref{fig:compare}(a) to be $\gamma_j=g(\omega_j)=2.8209\times
10^{12}/\sqrt{(\omega_{j}-\omega)^2+10^{12}}$Hz, and the flat
spectrum in Fig~\ref{fig:compare}(b) to be $\gamma_j=5.6419\times
10^{6}$ Hz with $j=1,2,\ldots 200$. For the flat spectrum we assume
the frequencies associated with the system and reservoir to be
$\omega=10^{9}$Hz and $\omega_j=j\times10^{7}$Hz $(j=1,2,\ldots
200)$ respectively. For the Lorentzian spectrum, however, we change
the frequencies of the reservoir to be
$\omega_j=(50+j/2)\times10^{7}$Hz $(j=1,2,\ldots 200)$ due to the
shape of the Lorentzian spectrum, which varies dramatically at the
center and is negligible at two sides. By making this change we can
sample the spectrum better. Then we plot Fig.~\ref{fig:compare}.
We can find that while the master equation leads to a good
approximate solution  in some cases, it fails sometimes. Thus our
method is more reliable, and the accuracy can be further improved by
using better numerical methods.

 We also note that the parity kicks can be done by
increasing the frequency of the harmonic oscillator for a short time
interval. For example, this can be achieved in the ion trap by
changing the electric field. (also see Refs.~\cite{squ1,squ2} for
schemes of generating squeezed states in ion trap)

{\it Conclusions.---} We have shown that for a general Hamiltonian
with bononic quadratic forms, we can compute dynamics of system
using exponential quadratic operators. Our method provides
substantial improvement over computation involving master equations
as we do not need to solve any differential equations and provides
numerical solution for Hamiltonians that can be written in quadratic
form of creation and annihilation operators. Thus, this new
techniques compares well with the dynamics of the system under a
master equation but it is in some sense more appealing as it could
provide in principle analytical expressions for some cases. In
particular, we analyze the effect of reservoir in a squeezing
process and we propose possible scheme to improve the degree of
squeezing. Our method can be applied to study the problem on the
quantization of nano-mechanical systems, the further work will be
presented elsewhere.

KLC acknowledges financial support by the National Research
Foundation \& Ministry of Education, Singapore. This work was
supported in part by the National Basic Research Program of China
grant No. 2007CB907900 and 2007CB807901, NSFC grant No. 60725416 and
China Hi-Tech program grant No. 2006AA01Z420.

\end{document}